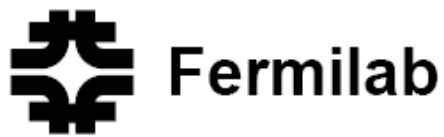

# BEAM-INDUCED EFFECTS AND RADIOLOGICAL ISSUES IN HIGH-INTENSITY HIGH-ENERGY FIXED TARGET EXPERIMENTS[*†]

N.V. Mokhov[#], S.R. Childress, A.I. Drozhdin,
V.S. Pronskikh, D. Reitzner, I.S. Tropin, K. Vaziri

Fermilab, Batavia, IL 60510, U.S.A.

## Abstract

The next generation of accelerators for Megawatt proton and heavy-ion beams moves us into a completely new domain of extreme specific energies of up to 0.1 MJ/g (Megajoule/gram) and specific power up to 1 TW/g (Terawatt/gram) in beam interactions with matter. This paper is focused on deleterious effects of controlled and uncontrolled impacts of high-intensity beams on components of beam-lines, target stations, beam absorbers, shielding and environment. Two new experiments at Fermilab are taken as an example. The Long-Baseline Neutrino Experiment (LBNE) will explore the interactions and transformations of the world's highest-intensity neutrino beam by sending it from Fermilab more than 1,000 kilometers through the Earth's mantle to a large liquid argon detector. The Mu2e experiment is devoted to studies of the conversion of a negative muon to electron in the field of a nucleus without emission of neutrinos.

---

[*]Work supported by Fermi Research Alliance, LLC under contract No. DE-AC02-07CH11359 with the U.S. Department of Energy.
[†]Presented paper at 12[th] International Conference on Radiation Shielding, September 2-7, 2012, Nara, Japan
[#]mokhov@fnal.gov

# Beam-Induced Effects and Radiological Issues in High-Intensity High-Energy Fixed Target Experiments

Nikolai Mokhov[*], Sam Childress, Alexandr Drozhdin,
Vitaly Pronskikh, Diane Reitzner, Igor Tropin, Kamran Vaziri

*Fermi National Accelerator Laboratory, Batavia IL 60510-5011, USA*

The next generation of accelerators for Megawatt proton and heavy-ion beams moves us into a completely new domain of extreme specific energies of up to 0.1 MJ/g (Megajoule/gram) and specific power up to 1 TW/g (Terawatt/gram) in beam interactions with matter. This paper is focused on deleterious effects of controlled and uncontrolled impacts of high-intensity beams on components of beam-lines, target stations, beam absorbers, shielding and environment. Two new experiments at Fermilab are taken as an example. The Long-Baseline Neutrino Experiment (LBNE) will explore the interactions and transformations of the world's highest-intensity neutrino beam by sending it from Fermilab more than 1,000 kilometers through the Earth's mantle to a large liquid argon detector. The Mu2e experiment is devoted to studies of the conversion of a negative muon to electron in the field of a nucleus without emission of neutrinos.

## 1. Introduction

Radiological issues and requirements are considered for related systems: shielding, groundwater and surface water, air emissions, residual activation of components, and prompt radiation. Consequences of accidental and operational losses of a 120-GeV proton beam in the LBNE [1] beam-line (**Figure 1**) are simulated with the MARS15 [2] and STRUCT [3] codes. The tolerable beam loss limits are derived with respect to the beam-line component integrity and impact on environment. Results of a comprehensive radiation study are presented for the LBNE target station, decay channel, hadron absorber system and the facility boundary. Parameters of the iron/concrete shielding, bulk soil shielding outside and site boundary arrangement with respect to the target station are derived. The muon range-out distance is also determined.

One of the main parts of the Mu2e [4] experimental setup is its superconducting production solenoid (PS), in which negative pions are generated in interactions of the primary proton beam with a target. The off-axis 8-GeV proton beam will deliver $6\times10^{12}$ protons/sec to the tungsten target, placed at the center of the PS bore. In order for the PS superconducting magnet to operate reliably, the peak neutron flux in the PS coils must be reduced by 3 orders of magnitude by means of a sophisticated absorber, optimized for the performance and cost. An issue with radiation damage is related to large residual electrical resistivity degradation in the superconducting coils, especially its Al stabilizer.

## 2. LBNE experiment

The neutrino beam will be generated using a proton beam in an energy range that maximizes the proton beam power delivered to the production target. The primary beamline is designed to transport high-intensity proton beam in the energy range of 60-120 GeV to the LBNE graphite target. The lifetime of the LBNE beamline facility is expected to be about 30 years. Some components are expected to last significantly longer so that they can meet radiological requirements until decommissioning.

### 2.1. Beam loss in primary beamline

The main criteria which have guided design of the primary beam line is transmission of high intensity beam with minimum losses and precision of targeting, keeping activation of components and ground water below the regulatory limits at normal and accidental conditions.

*Corresponding author. Email: mokhov@fnal.gov

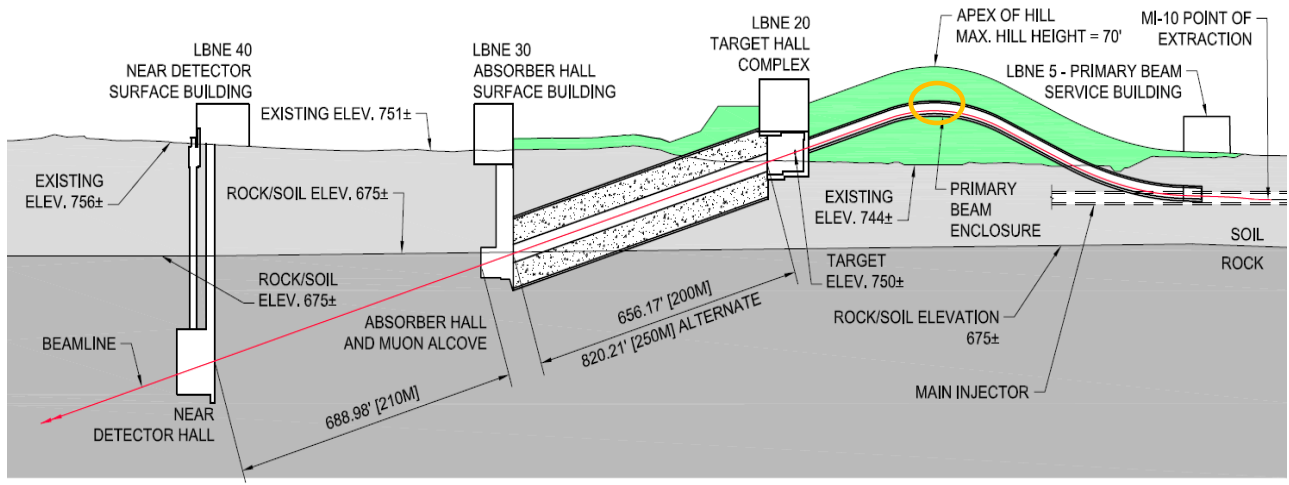

Figure 1. LBNE target hall and beamline shielding. Elevations and height are shown in feet. Lengths of decay channel and dirt shielding between muon alcove and near detector hall are shown in feet and meters. Shown in yellow is the region of the highest beam loss rate and, respectively, of the highest radiation levels.

STRUCT and MARS simulations have evaluated the impact of a localized full beam loss at any location along the beamline and a sustained small fractional loss. The proton beam energy considered in these studies was 120 GeV. In the first case, peak beam pipe temperature of twice the melting point for stainless steel is reached with a single lost full beam pulse of $1.6 \times 10^{14}$ ppp. At initial intensity of $4.9 \times 10^{13}$ ppp, beam pipe failure is probable after 4-5 lost full beam pulses. Therefore, large beam loss for even a single pulse needs to be robustly prevented via an Integrated Beam Permit System.

In the second case, magnitude of the beam loss is chosen as a value which is within accuracy limitations for intensity monitors at beginning and end of the beamline, and which might not produce a vacuum failure. The STRUCT simulations have shown that the highest loss rate takes place in the quadrupole magnet and two adjacent dipole magnets located right at the apex of the primary beam line (**Figure 1**). Under this condition, for a scenario where there is accidental beam loss at 0.3% of the beam for 30 continuous days, calculated with MARS15 the peak contact dose after 24 hours of cool-down is: 1) for tunnel walls, 5 mSv/hr over a 20-ft (~6m) region of the tunnel, and > 1 mSv/hr over a 50-ft (~15m) tunnel region, 2) for the hottest magnet, 500 mSv/hr over ~1m of magnet steel, and > 100 mSv/hr over most of a 3-m magnet.

Even after waiting 6 months with no beam, a magnet would still be at > 30 mSv/hr in the hottest region. The Fermilab limit of "0.5 mSv/hr on contact to safely permit all necessary maintenance" dictates sustained localized beam loss to be a factor of one thousand less than considered above, in a good agreement with NuMI requirements [5]. The Integrated Beam Permit System will again take care of this.

### 2.2. Radiological requirements and shielding

Radiological requirements for the design of the beam lines and experimental facilities are described in the Fermilab Radiological Control Manual [6]. The 2.3 MW LBNE proton beam is produced in the form of $1.6 \times 10^{14}$ protons every 1.333 seconds. Based on the experience with operation of the NuMI beam line, it was concluded that for the radiological calculation, a continuous beam loss rate of $10^{-5}$ and an accidental loss of 2 pulses per hour will be used.

The beamline passes through the aquifer regions, therefore radiation requirements are quite stringent and vary from region to region. The calculated soil shielding required for 2.3 MW beam, for unlimited occupancy classification (1 mSv/year), is 6.4 m for continuous fractional beam loss of $10^{-5}$ level and 7 m for two localized full beam pulses lost per hour. To reduce the accidental muon dose at the site boundary to < 10 μSv/year, 122 m of soil path of the muons is required.

### 2.3. Prompt dose

Prompt radiation is one of the main issues in the above-grade target option. There are two contributions to prompt dose rate at both onsite and offsite locations: direct radiation outside the shielding and sky-shine, which is primarily neutron back-scattered radiation from air. The primary beam transport line, target hall and the decay pipe (**Figure 2**), as sources, contribute to dose in areas accessible to members of the public. The LBNE beam line is pointed towards the nearest site boundary. Therefore, the direct muon dose adds up to the prompt doses from other sources, at the nearest site boundary. Based on the MARS calculations, both the annual direct and sky-shine doses are calculated for both offsite and onsite locations. Direct accidental muon dose at the apex of the transport line is also included in the offsite dose (**Table 1**).

Nearest site boundary from the target hall and decay pipe is 520 meters away. Wilson Hall is the nearest routinely occupied on-site building, which is 1050 meters away. To allow operations of other experiments, beamlines and accelerators, the offsite goal for LBNE is set at 20 μSv/year, from all radiation sources generated by this beamline. The total offsite dose, at the nearest site boundary, due to both direct and sky-shine is estimated to be 13.2 μSv/year. Both the maximum direct and sky-shine annual dose to the occupants of the Wilson Hall has been calculated. The total annual dose, at Wilson Hall, due to both direct and sky-shine is estimated to be 0.6 μSv. Doses for other locations onsite, which are further away, will be less.

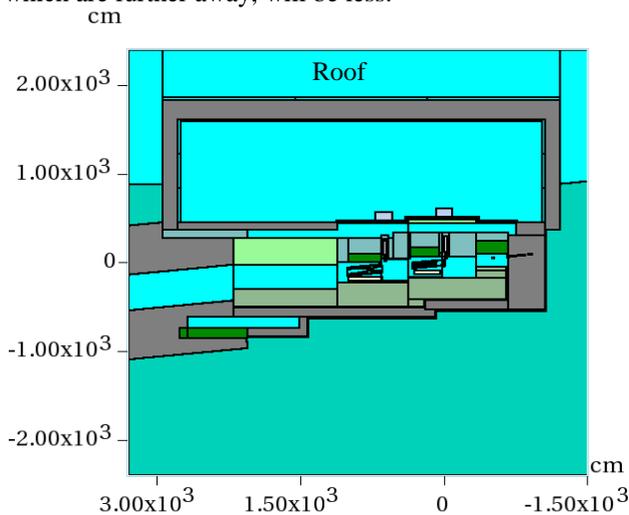

Figure 2. MARS15 model of LBNE target hall. Concrete roof with vertical concrete walls are shown.

Table 1. Contributions to the total offsite and onsite doses doses for the main LBNE regions (see **Figures 1** and **2**).

| LBNE regions | Site Boundary (mSv/year) | Wilson Hall (mSv/year) |
|---|---|---|
| Decay pipe above ground section, sky-shine | 2.155 | 0.138 |
| Decay pipe above ground section, direct | 0.100 | 0.000 |
| Target hall with concrete roof, sky-shine | 5.844 | 0.375 |
| Target hall walls, direct | 1.835 | 0.006 |
| Target hall roof, direct | 0.037 | 0.070 |
| Absorber hall service building roof, sky-shine | 2.800 | 0.017 |
| Absorber hall service building roof, direct | 0.007 | 0.001 |
| Primary transport line, sky-shine | 0.066 | 0.004 |
| Transport line muons, direct | 0.333 | 0.000 |
| **Total, sky-shine** | **10.9** | **0.5** |
| **Total annual dose, direct and sky-shine** | **13.2** | **0.6** |

## 2.4. Surface and groundwater contamination

Production of potentially mobile isotopes such as $^3$H and $^{22}$Na is an unavoidable consequence of high-energy particle collisions with nuclei. Since all of the Mu2e beam line and the primary transport line of LBNE are located in the glacial till, with no direct connection to the aquifer, all radionuclides produced in the soil below the enclosures will have to move down through the different soil layers to reach the aquifer. The seepage velocities, for the layers in the glacial till, are very small and the concentrations of the radionuclides are reduced by 5 to 7 orders of magnitude.

The LBNE target hall shielding is designed such that to render the cumulative concentration of the radionuclides, outside the protective layers, over 30 years to be less than 30% of the regulatory surface limit. At the same time, the decay pipe and the absorber hall are designed with sufficient shielding and water impermeable layers to render the cumulative concentration of the radionuclides in the soil over 30 years to be less than the current standard detection limits. The current accepted detection limits are 1 pCi/ml (37 Bq/l) for $^3$H and 0.04pCi/ml (1.48 Bq/l) for $^{22}$Na. Due to the proximity of these areas to the aquifer, the stricter concentration limit is used.

## 2.5. Air activation

High levels of radioactive air are produced in the target area and the decay pipe. The air to these two locations is in a closed and isolated loop. The two separate streams of air that feed the target chase and the decay pipe are combined on return and sent to the air handling room. In the air handling room, this air is chilled and dehumidified before returning to the decay pipe and the target chase. The air handling room structure and the doors are designed to be air tight. A few percent of the target chase air leaks into the target hall, where it is dehumidified and sent to the downstream of the beam transport line enclosure. Radionuclides from air include $^{16}$N ($t_{1/2}$ = 7sec), $^{14}$O ($t_{1/2}$ = 70sec), $^{15}$O ($t_{1/2}$ = 2min), $^{13}$N ($t_{1/2}$ = 10min), $^{11}$C ($t_{1/2}$ = 20min), and $^{41}$Ar ($t_{1/2}$ = 110min). For the target hall design ventilation rate, the air transit/decay time is about 3.5 hours before release. Similar, long-transit-time method is used to reduce the activation levels of the air releases from the Mu2e target hall. Radioactive air from the target hall and the beam line flows back all the way upstream in the primary beamline enclosure and released. The transit time from LBNE and Mu2e target halls to upstream end of the transport line exhaust is sufficiently long to allow the airborne radionuclides to decay by several orders of magnitude.

Current Fermilab radioactive air emission permit allows the annual exposure of a member of public offsite to the radioactive air emissions, from all sources to be less than 1 μSv. It is the goal of the LBNE and Mu2e designs to have the air emissions to contribute to less than 30%-50% of this limit, leaving room for emissions from other accelerators and beam lines at the laboratory.

## 3. Mu2e experiment

The Mu2e experiment [4] is devoted to studies of the charged lepton flavor violation which up to now has never been observed and can manifest itself as the conversion of $\mu^-$ to $e^-$ in the field of a nucleus without emission of neutrinos.

### 3.1. Production Solenoid design

One of the main parts of the Mu2e experimental setup is its production solenoid (PS), in which negative pions are generated in interactions of the primary proton beam with the target (see **Figure 3**). These pions then decay into muons which are delivered by the transport solenoid to the detectors. The off-axis 8 GeV proton beam will deliver $6\times10^{12}$ protons per second to the heavy metal target, placed at the center of the PS bore.

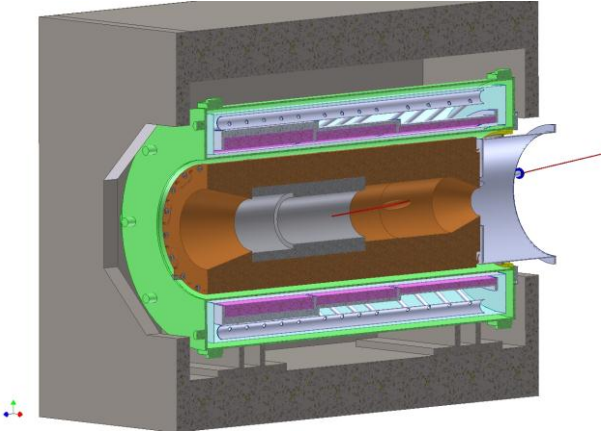

Figure 3. Mu2e production solenoid with bronze absorber.

The constraints in the PS shielding insert (absorber) design are quench stability of the superconducting coils, low dynamic heat loads to the cryogenic system, a reasonable lifetime of the coil components, acceptable hands-on maintenance conditions, compactness of the absorber that should fit into the PS bore and an aperture large enough to not compromise pion collection efficiency, cost, weight and engineering requirements.

Thorough optimization of the absorber design was performed with MARS15. The following quantities were focused on: dynamic heat load, peak power density, number of displacements per atom (DPA) in the helium-cooled solenoid coils, peak absorbed dose and peak neutron flux in the coils. As an example, neutron flux isocontours are shown in **Figure 4**.

Limits on the radiation quantities were set [7] based on: quench protection requiring that peak coil temperature does not violate allowable value of 5 K with 1.5 K thermal margin for peak power density, 10% degradation of ultimate tensile strength for absorbed dose, RRR (residual resistivity ratio) degradation from ~1000 to ~100 in Al stabilizer for DPA, and requirements from the particular cooling system designed for dynamic heat load. In the current design all the quantities analyzed, peak power density 18 (30) µW/g, peak DPA/year 4 (6)$10^{-5}$, peak absorbed dose/lifetime 1.7 (7) MGy, and dynamic heat load 20 (100) W satisfy the limits shown in parentheses. The annual peak DPA rate along the PS inner coil is shown in **Figure 5**.

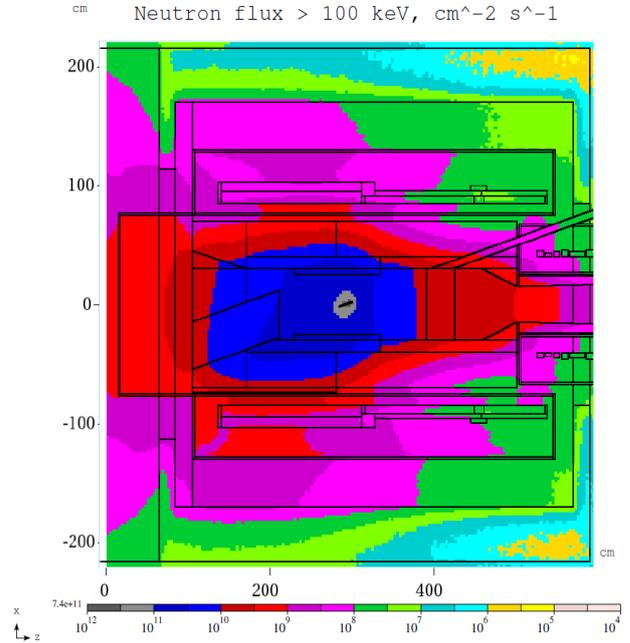

Figure 4. Neutron flux in Mu2e production solenoid.

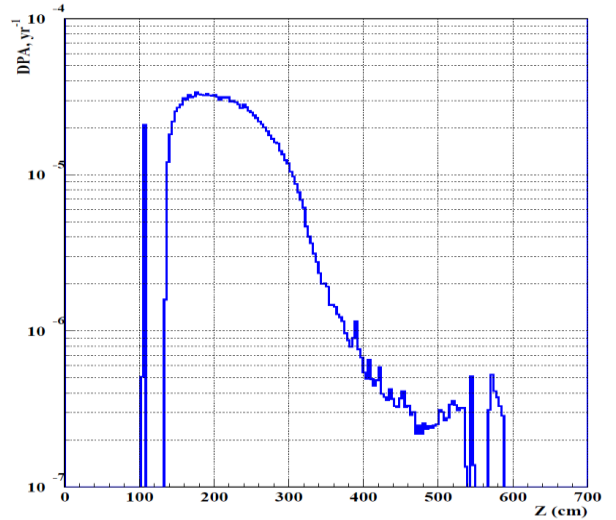

Figure 5. Longitudinal distribution of peak DPA rate in PS inner coil.

The 3D thermal analysis was performed for the radiation heat load in case of the optimized absorber design. The dynamic heat load map in the coil and the support structure generated by the MARS15 code was applied to all parts of the cold mass. The Final Element model created by COMSOL Multiphysics [8] was discretized to the level of individual layers and the interlayer insulation/conducting sheets. The maximum temperature of 4.8 K is found in the middle of the inner surface of the thickest coil section (**Figure 6**).

Decay heat of the target was calculated using the DeTra mode [9, 10] in MARS15. **Figure 7** shows the contribution of 31 major (60%) nuclides to decay heat in the target. The total decay heat after one year of irradiation is estimated to be 11.3 W, much lower compared to the dynamic heat load of 800 W.

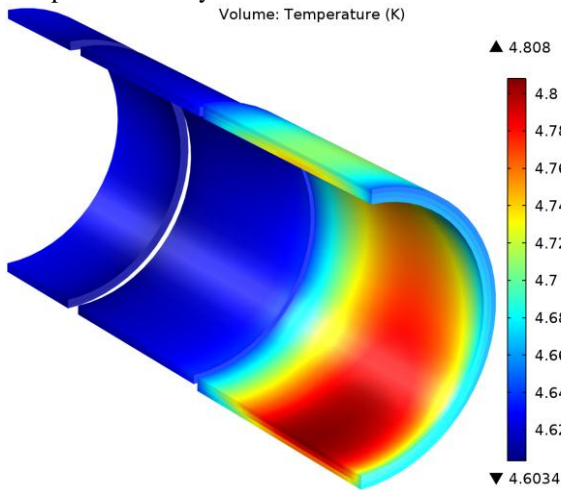

Figure 6. Temperature distribution in Mu2e inner coil.

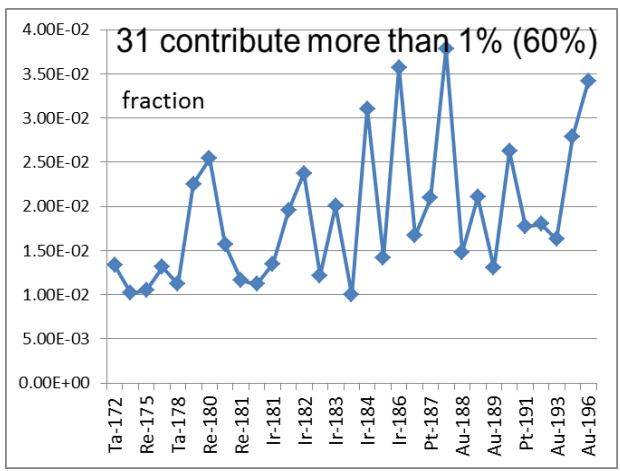

Figure 7. Relative contribution of 31 major nuclides to decay heat of Mu2e target. These contribute more than 1% each resulting in 60% of the total decay heat.

Residual dose on contact with PS coils and Al stabilizer is ~0.7 mSv/hr after one year of irradiation and one week of cooling, and ~80 μSv/hr after 30 days of irradiation and one week of cooling, which is rather high and requires particular safety measures for personnel performing PS maintenance during the lifetime of the experiment.

*3.2. Surface, groundwater and air contamination*

Based on MARS15 simulations of the hadron flux and star density and using the Fermilab standard Concentration Model [6] at the design intensity, the average concentrations of radionuclides in the sump pump discharge were calculated as 24 pCi/ml (888 Bq/l) due to $^3$H and 2 pCi/ml (74 Bq/l) due to $^{22}$Na. This is 2% of the total surface water limit if the pumping is performed once a month (conservative scenario). Build-up of these nuclides in ground water at $1.2 \times 10^{20}$ protons per year will be as low as $6.2 \times 10^{-8}$ % of the total limit over three years of operation.

Air activation and flow estimations show that at 500 cfm or 0.236 m$^3$/s, for the configuration without a pipe connecting the target region to the beam dump (average hadron flux over the whole hall volume is $5.5 \times 10^6$ cm$^{-2}$s$^{-1}$), the maximum annual activity released from the target hall is less than 29 Ci ($1.073 \times 10^{12}$ Bq), which is ~ 7% of the Lab's air release limit.

**4. Conclusion**

High intensity, high energy beams required for the new generation of the high energy experiments provide new challenges to the material and shielding aspects of the design of these experiments. However, careful design with good reliance on the Monte Carlo simulations on every step of the design makes constructing LBNE and Mu2e beam lines and experiments manageable.

**Acknowledgements**

Work supported by Fermi Research Alliance, LLC, under contract DE-AC02-07CH11359 with the U.S. Department of Energy.